# Dynamo models and differential rotation in late-type rapidly rotating stars

Eurico Covas[*,1], David Moss[**,2], and Reza Tavakol[***,1]

[1] Astronomy Unit, School of Mathematical Sciences, Queen Mary, University of London, Mile End Road, London E1 4NS, UK
[2] Department of Mathematics, The University, Manchester M13 9PL, UK

**Abstract.** Increasing evidence is becoming available about not only the surface differential rotation of rapidly rotating cool stars but, in a small number of cases, also about temporal variations, which possibly are analogous to the solar torsional oscillations. Given the present difficulties in resolving the precise nature of such variations, due to both the short length and poor resolution of the available data, theoretical input is vital to help assess the modes of behaviour that might be expected, and will facilitate interpretation of the observations. Here we take a first step in this direction by studying the variations in the convection zones of such stars, using a two dimensional axisymmetric mean field dynamo model operating in a spherical shell in which the only nonlinearity is the action of the azimuthal component of the Lorentz force of the dynamo generated magnetic field on the stellar angular velocity. We consider three families of models with different depths of dynamo-active regions. For moderately supercritical dynamo numbers we find torsional oscillations that penetrate all the way down to the bottom of the convection zones, similar to the case of the Sun. For larger dynamo numbers we find fragmentation in some cases and sometimes there are other dynamical modes of behaviour, including quasi-periodicity and chaos. We find that the largest deviations in the angular velocity distribution caused by the Lorentz force are of the order of few percent, implying that the original assumed 'background' rotation field is not strongly distorted.

**Key words.** Stars: magnetic fields – torsional oscillations – activity – rapidly rotating cool stars

## 1. Introduction

A steadily increasing number of observations of the variations in the surface rotation of rapidly rotating cool stars is now becoming available (e.g. Collier Cameron & Donati 2002; Donati, Collier Cameron & Petit 2003; Reiners & Schmitt 2003; Messina & Guinan 2003, amongst other papers too numerous to mention here). Using a variety of techniques including spot tracing and Doppler imaging, these observations parallel, at much lower resolution, determinations of variations in the solar surface rotation law.

For example, recent observations of the rapidly rotating K dwarf AB Doradus ($P_{\rm equator} \approx 0.5$d), and of other rapidly rotating late-type stars, show the presence of a surface differential rotation generically similar to that of the Sun, in that there is equatorial acceleration, with a mean pole-equator variation of about 0.5%. In one way the differential rotation of AB Dor is remarkably solar-like in spite of the very different rotational period, in that the equator pulls one full turn ahead of the pole every 110 days or so, compared to a lap time of 120 days in the case of the Sun. This differential rotation varies by about a factor of 2 (Collier, Cameron & Donati 2002; Donati *et al.* 2003b).

LQ Hydrae is another well studied rapidly rotating dwarf, with $P_{\rm equator} \approx 1.6$d. Again there is equatorial acceleration, with a mean latitudinal differential rotation of about 1%, but varying with time by a factor of about 10 (Donati *et al.* 2003b).

A significant trend seems to be that surface differential rotation and rotational period are positively correlated, so that the product $\Omega\Delta\Omega$ does not depend strongly on rotational period (e.g. Hall 1991; Donahue 1993; Messina & Guinan 2003; Reiners & Schmitt 2003), although it is clear from the above that there are exceptions.

In general, these rapid rotators appear to possess magnetic cycles (e.g. Donati et al. 2003a), in common with many of the cool stars in the Mount Wilson HK-project (e.g. Baliunas et al. 1995).

These observations raise a number of interesting questions, including the form of the surface rotation law and its variations. Of course, rotation laws and their variations in the interiors of these stars are quite unknown. Even at the surface, spatiotemporal variations are only beginning to be observed

*Send offprint requests to*: R. Tavakol
 [*] e-mail: e.o.covas@qmul.ac.uk
 [**] e-mail: moss@ma.man.ac.uk
 [***] e-mail: r.tavakol@qmul.ac.uk

for a few objects. In particular, it is of considerable interest whether torsional oscillations of the type observed in the Sun are widespread in these stars – the observations of LQ Hyd and AB Dor give tantalising hints.

Here we examine these questions in the context of nonlinear mean field dynamo models in a spherical shell, which include the feedback of the Lorentz force of the large-scale magnetic field on this rotation law as the nonlinearity. This parallels our previous studies of solar torsional oscillations (see, e.g., Covas *et al.* 2000; Tavakol *et al.* 2002; Covas *et al.* 2004 and references therein). We investigate the nature of variations in the convection zones of rapidly rotating stars, by considering three families of models with different depths of dynamo-active regions ('convective envelopes'). Specialisation to rapid rotation is made by choice of a quasi-cylindrical zero order rotation law. This is subsequently modified as the dynamo saturates, via the feedback of the azimuthal Lorentz force. We study the effects of changing the strength of the zeroth order differential rotation, and of the $\alpha$–effect on the variations caused to the zero order rotation profile. By examining this range of models we aim to obtain insight into the possible forms and types of variation in their rotation that might be expected. Such theoretical input is, despite its obvious limitations, potentially of primary importance given the scarcity and limited duration and resolution of observations of rotational variations of these stars.

Furthermore, theoretical stellar dynamo studies are in a fundamentally unsatisfactory state because there has only been one object, the Sun, against which to calibrate in anything but the grossest manner the unknown parameters occurring in mean field dynamo models. Modelling late-type stars with known surface rotation laws (and in a few cases, known temporal variations) has the potential to calibrate and test dynamo theory models and predictions against a range of observed systems.

It can be noted that these rapidly rotating stars with significant convective envelopes often display nonaxisymmetric features, such as photometric variability believed to be caused by large starspots. These features are conventionally associated with the presence of large-scale nonaxisymmetric magnetic fields. Here we only study axisymmetric fields as a first approximation, taking the viewpoint that the nonaxisymmetric components can be treated as higher order contributions to the field. Moss (2004a,b) has made a preliminary study of nonaxisymmetric field generation in these object, using a different (alpha-quenching) dynamo nonlinearity.

In the next section we outline our dynamo model. Sect. 3 contains our detailed results. In Sect. 4 we make a brief comparison of the results of our model with some observations and in Sect. 5 we summarise our results and draw some conclusions.

## 2. The model

We shall assume that the gross features of the large-scale magnetic field in such stars can be described by a mean field dynamo model, with the standard equation

$$\frac{\partial \boldsymbol{B}}{\partial t} = \nabla \times (\boldsymbol{u} \times \boldsymbol{B} + \alpha \boldsymbol{B} - \eta \nabla \times \boldsymbol{B}). \tag{1}$$

Here $\boldsymbol{u} = v\hat{\boldsymbol{\phi}} - \frac{1}{2}\nabla\eta$, the term proportional to $\nabla\eta$ represents the effects of turbulent diamagnetism, and the velocity field is taken to be of the form $v = v_0 + v'$, where $v_0 = \Omega_0 r \sin\theta$, $\Omega_0$ is a prescribed underlying rotation law and the component $v'$ satisfies

$$\frac{\partial v'}{\partial t} = \frac{(\nabla \times \boldsymbol{B}) \times \boldsymbol{B}}{\mu_0 \rho r \sin\theta} . \hat{\boldsymbol{\phi}} + \nu D^2 v', \tag{2}$$

where $D^2$ is the operator $\frac{\partial^2}{\partial r^2} + \frac{2}{r}\frac{\partial}{\partial r} + \frac{1}{r^2 \sin\theta}(\frac{\partial}{\partial \theta}(\sin\theta\frac{\partial}{\partial \theta}) - \frac{1}{\sin\theta})$ and $\mu_0$ is the induction constant. The assumption of axisymmetry allows the field $\boldsymbol{B}$ to be split simply into toroidal and poloidal parts, $\boldsymbol{B} = \boldsymbol{B}_T + \boldsymbol{B}_P = B\hat{\boldsymbol{\phi}} + \nabla \times A\hat{\boldsymbol{\phi}}$, and equation (1) then yields two scalar equations for $A$ and $B$. Nondimensionalizing in terms of the stellar radius $R$ and time $R^2/\eta_0$, where $\eta_0$ is the maximum value of $\eta$, and putting $\Omega = \Omega^*\tilde{\Omega}$, $\alpha = \alpha_0\tilde{\alpha}$, $\eta = \eta_0\tilde{\eta}$, $\boldsymbol{B} = B_0\tilde{\boldsymbol{B}}$ and $v' = \Omega^*R\tilde{v}'$, results in a system of equations for $A, B$ and $v'$. The dynamo parameters are $R_\alpha = \alpha_0 R/\eta_0$, $R_\omega = \Omega^* R^2/\eta_0$, $P_r = \nu_0/\eta_0$, and $\tilde{\eta} = \eta/\eta_0$, where $\Omega^*$ is the stellar surface equatorial angular velocity. Here $\eta_0$ and $\nu_0$ are the turbulent magnetic diffusivity and viscosity respectively and $P_r$ is the turbulent Prandtl number. From here onwards we will refer to dimensionless quantities, unless explicitly stated otherwise, dropping the tildes.

Equations (1) and (2) were solved using the code described in Moss & Brooke (2000) (see also Covas *et al.* 2000) together with the boundary conditions given below, over the range $r_0 \leq r \leq 1$, $0 \leq \theta \leq \pi$. We consider three families of models with dynamo-active shells ('convection zones') of different thicknesses with lower boundaries at fractional radii $r_0 = 0.2, 0.64$ and $0.775$. We associate these with physical "stellar radii" $R = 5.6 \times 10^{10}$ cm, $6.96 \times 10^{10}$ cm and $8 \times 10^{10}$ cm respectively. The stellar convection zone proper is thought to occupy the region with fractional radius $r \gtrsim r_{tc}$, where $r_{tc} = 0.25, 0.7, 0.8$ respectively and where the regions $r_0 \leq r \lesssim r_{tc} \pm d$, where $d = 0.05$, can be thought of as the corresponding overshoot regions/tachoclines. In the simulations discussed below we use a mesh resolution of $61 \times 101$ points, uniformly distributed in radius and latitude respectively.

In order to represent the expected form of the rotation law in rapidly rotating stars with Rossby number $Ro \gg 1$, in $r_0 \leq r \leq 1$ we chose

$$\Omega_0(r, \theta) = \left[\frac{\frac{1}{2}\{1 + \text{erf}[(r - r_{tc})/d]\} ar^2 \sin^2(\theta) + 1}{(1 + a)}\right]\Omega^* \tag{3}$$

for our zero-order rotation law, where $a$ controls the strength of the differential rotation and $\Omega^*$ is a constant. Contours $\Omega_0 =$ constant, with $r_0 = 0.64$, are shown in Fig 1.

This choice is consistent also with the limited observations, which suggest that mean surface rotation laws are of the solar-like form $\Omega_s = c + d \sin^2\theta$, where $c$ and $d$ are constants and $\theta$ is the co-latitude. This form of $\Omega_0(r, \theta)$ means that the relevant dynamo number is actually

$$\frac{a}{1+a} R_\omega = R_\omega^*, \tag{4}$$

say, and that $a$ and $R_\omega$ are not independent parameters. We adopt the strategy of fixing $R_\omega$ for each value of $r_0$, and allowing $a$, and so $R_\omega^*$, to vary.

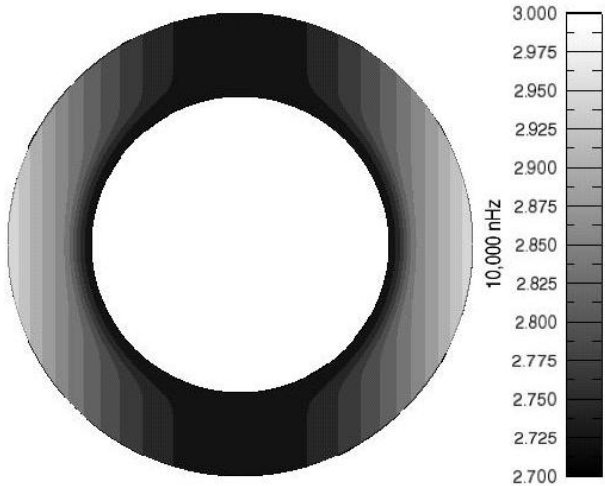

**Fig. 1.** Contours of the zero-order rotation profile $\Omega_0$ = constant with $r_0 = 0.64$. The darker contours indicate slower rotation.

If, to be definite, we choose $R_\omega = 10^5$ when $\Omega = 10\Omega_\odot$, $R = R_\odot$, then we obtain $\eta_0 \approx 1.45 \times 10^{12} \text{cm}^2\text{s}^{-1}$. Then, e.g., calculations with $\Omega = 10\Omega_\odot$, $a = 0.1$ and $\Omega = 19\Omega_\odot$ (cf. LQ Hydrae), $a = 0.05$ both have $R_\omega^* = 9.1 \times 10^3$ and so will yield identical results in terms of dimensionless variables. The basic arbitrariness in our nominal value of $\eta_0$ should be remembered, especially when comparing our results with observations (Sect. 4). $R_\alpha$ is rather difficult to estimate, but values $1 \lesssim R_\alpha \lesssim 100$ seem acceptable.

Rather arbitrarily, we put $\alpha(r, \theta) = \alpha_r(r) \cos(\theta)$, where

$$\alpha_r = \begin{cases} 0 & if \ r < r_1 \\ 1 & if \ r > r_2 \\ \left[\frac{(r-r_1)}{(r_2-r_1)}\right]^2 \left[3 - 2\frac{(r-r_1)}{(r_2-r_1)}\right] & \text{elsewhere.} \end{cases}$$

The parameters $r_1$ and $r_2$ used in our models are given in Table 1.

**Table 1.** Values of model parameters as $r_0$ varies.

| $r_0$ | $r_1$ | $r_2$ | $r_3$ | $r_4$ | $R$(cm) | $R_\omega$ |
|---|---|---|---|---|---|---|
| 0.20 | 0.25 | 0.40 | 0.25 | 0.35 | $5.6 \times 10^{10}$ | $6.49 \times 10^4$ |
| 0.64 | 0.70 | 0.80 | 0.70 | 0.80 | $6.96 \times 10^{10}$ | $10^5$ |
| 0.775 | 0.80 | 0.85 | 0.80 | 0.85 | $8.0 \times 10^{10}$ | $1.32 \times 10^5$ |

Also, in order to take some notional account of the likely decrease in the turbulent diffusion coefficient $\eta$ in the overshoot region, we took

$$\eta(r) = \begin{cases} \eta_{\min} & if \ r < r_3 \\ 1 & if \ r > r_4 \\ \eta_{\min} + \frac{(r-r_3)}{(r_4-r_3)}(1 - \eta_{\min}) & \text{elsewhere.} \end{cases}$$

The values of the parameters $r_3$ and $r_4$ are also given in Table 1.

Throughout we take $\alpha_r > 0$ and $R_\alpha < 0$ and use a uniform density: our earlier work (Covas et al. 2004) illustrates the effects of introducing a strongly radially dependent density – in brief, no qualitatively new effects are found, although the radial distribution of perturbations to the angular velocity can be altered. In particular, it tends to increase (decrease) the strength of the torsional oscillations at the top (bottom) of the convection zone. At the outer boundary $r = 1$ we impose vacuum boundary conditions, whereby the interior poloidal field is smoothly joined, by a matrix multiplication, to an external vacuum solution; the azimuthal field there satisfies $B = 0$. At the inner boundary $r = r_0$ we use the same conditions as Tavakol et al. (2002). The variable $v'$ satisfies stress-free boundary conditions.

Our dynamo model shares features with both 'interface' and 'distributed' models – the radial rotational shear is concentrated near a 'tachocline', whereas the alpha effect is more radially homogeneous. We adopted this model here partly because of the success of a similar model in studying solar torsional oscillations (Covas et al. 2004 and references therein), and also because of the repeated hints from observations that a distributed dynamo may be more appropriate for these stars (e.g. Donati et al. 2003a,b, Petit et al. 2004). Meridional circulation has been suggested to play a role in the solar dynamo. We did not include it in our model, choosing in this exploratory study to direct our attention to a simple, quasi-interface, dynamo model. An investigation of the effects of meridional circulation on torsional oscillations would certainly be welcome. However we note that, even for the Sun, the circulation in the deeper convection zone can only be estimated by plausible extrapolation from that of the near-surface regions. The situation with rapid rotators is more uncertain, and an investigation of such a dynamo model would entail considerable exploration of this additional degree of freedom.

## 3. Results

For each value of $r_0$ we assign a "stellar radius", given above. Using $\eta_0$, we can then evaluate $R_\omega$ – see Table 1. We chose nominal values $\Omega^* = 3 \times 10^{-5} \text{rads}^{-1} \approx 10\Omega_\odot$, and $\eta_0 = 1.45 \times 10^{12} \text{cm}^2\text{s}^{-1}$ for all models. For each of these choices, we made a detailed study of dynamo action and the associated variations in the rotation laws as the strength of the alpha-effect and differential rotation (measured by $R_\alpha$ and $a$ or $R_\omega^*$ respectively) was varied.

Here we shall briefly discuss the different families of models separately:

### 3.1. Stars with deep convection zones: $r_0 = 0.2$

With differential rotation given by $a = 0.1$ ($R_\omega^* = 5.9 \times 10^3$), the onset of dynamo action occurs at $R_\alpha \approx -1.0$. The strongest torsional oscillations are found to be around the bottom of the dynamo region. As an example we show in Fig. 2 the radial ($r - t$) contours of the angular velocity residuals $\delta\Omega$ as a function of time. Also shown is the corresponding butterfly diagram (i.e. latitude-time plot) for the toroidal component of the magnetic field $\boldsymbol{B}$ near the surface and the butterfly diagram of the torsional oscillations at surface.

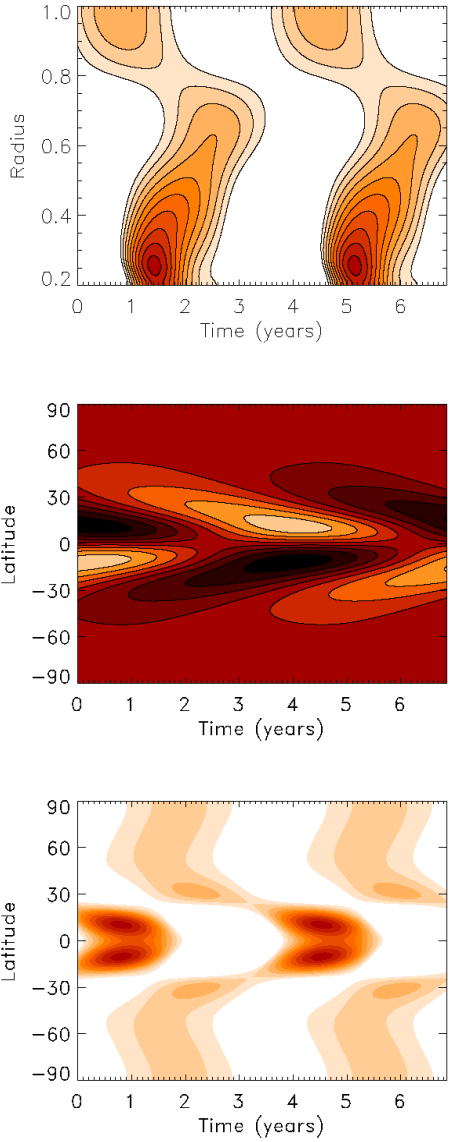

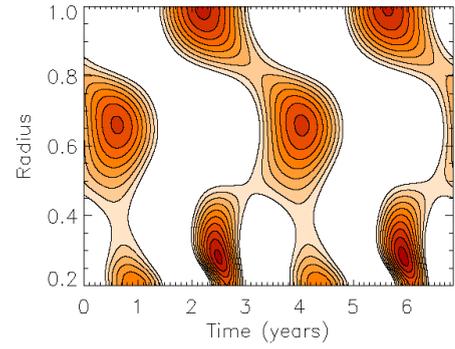

**Fig. 3.** The $r-t$ diagram for torsional oscillations showing spatiotemporal fragmentation. Note the contrast between the top of the dynamo region and the bottom, where the period is halved.

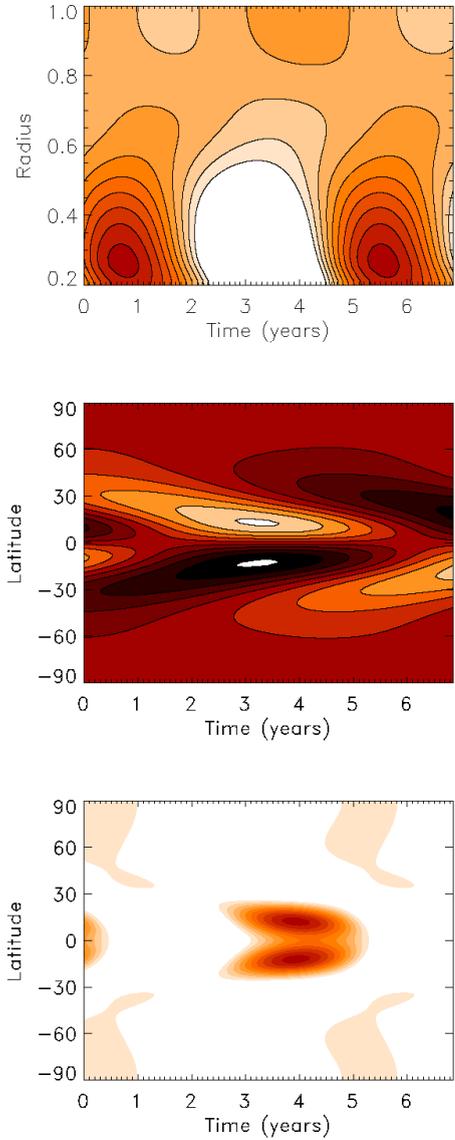

**Fig. 2.** The radial $r-t$ contours of the angular velocity residuals $\delta\Omega$ as a function of time for a cut at 10 degrees latitude, as well as the corresponding butterfly diagram (i.e. latitude-time plot) for the toroidal component of the magnetic field **B** and the torsional oscillation butterfly diagram near the surface for a dynamo region with $r_0 = 0.2$, $a = 0.1$ and $R_\alpha = -1.4$. Note the penetration of the torsional oscillations all the way down to the bottom of the dynamo region, and the presence of a weak polar branch.

For $R_\alpha$ between $-1.0$ and $-2.0$ we obtain torsional oscillations that extend all the way down to the bottom of the dynamo region. For $R_\alpha$ between about $-2.0$ and $-2.7$ there is some fragmentation, in the sense that the regimes at the top and bottom of the convection zone are not the same and in particular the latter has half the period of the former. An example of such fragmentation can be seen in Fig. 3. For still larger values of $R_\alpha$ the regime becomes noisy and then irregular.

When the differential parameter is given by $a = 0.01$ ($R_\omega^* = 6.43 \times 10^3$), the marginal value of $R_\alpha \approx -8.0$, with the strongest torsional oscillations again being found around the bottom of the dynamo region. As an example we show in Fig. 4 analogous results to Fig. 2 for this case.

**Fig. 4.** As in Fig. 2 for a dynamo region with $r_0 = 0.2$, $a = 0.01$ and $R_\alpha = -8$. Note the presence of torsional oscillations near both top and the bottom of the dynamo region.

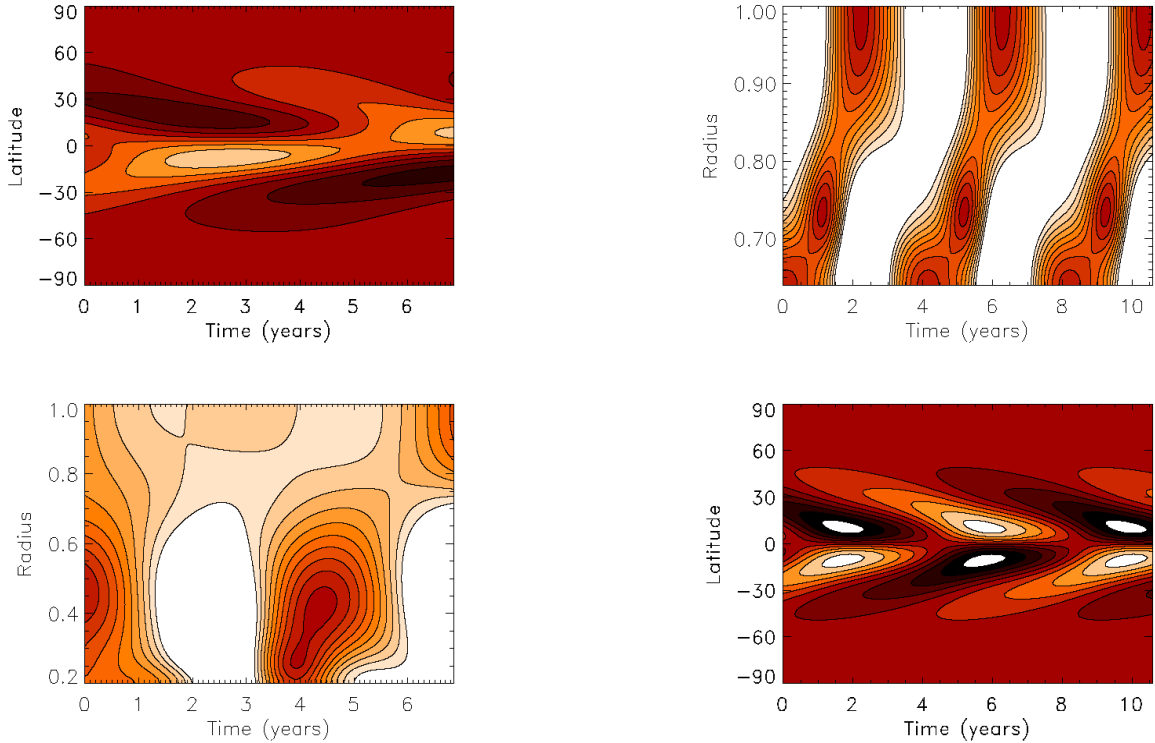

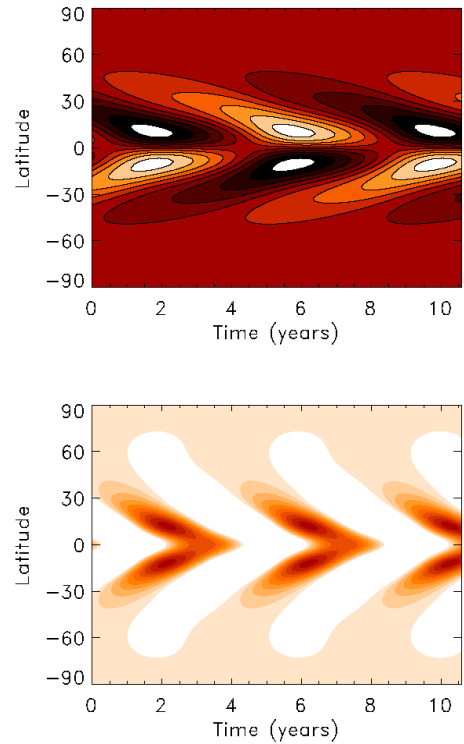

**Fig. 5.** An example of a noisy periodic regime before the onset of fully chaotic behaviour. The top panel shows the magnetic butterfly diagram close to the surface with the usual solar-type migration towards the equator, whilst also showing slight equatorial asymmetry. The bottom panel shows the $r - t$ diagram for the torsional oscillations, again showing the onset of break-up of periodicity.

Around $R_\alpha \sim -10$ the regime becomes quasi-periodic/noisily periodic with the butterfly diagram showing slight equatorial asymmetry and the torsional oscillations starting to break up, as shown in Fig. 5. For $R_\alpha$ between $\sim -12$ and $-14$ the angular velocity and magnetic fields still show oscillations but the migration towards the equator is lost.

### 3.2. Stars with intermediate/solar-type convection zones: $r_0 = 0.64$

With differential rotation defined by $a = 0.1$ ($R_\omega^* = 9.1 \times 10^3$), the marginal value of $R_\alpha$ is about $-2.0$. Solutions become noisily periodic for $R_\alpha \lesssim -3$.

With $a = 0.01$, the onset of dynamo action is at $R_\alpha \approx -10.0$, with the regime becoming quasi-periodic/noisily periodic around $R_\alpha \sim -14$.

As examples we show in Figs. 6 and 7 results analogous to those shown above, when $a = 0.1$ and $a = 0.01$ respectively.

**Fig. 6.** As in Fig. 2 for a dynamo region with $r_0 = 0.64$, $a = 0.1$ and $R_\alpha = -2$. Note the penetration of the torsional oscillations all the way down to the bottom of the dynamo region, and also the localisation of the torsional oscillations and the magnetic field near the equator.

### 3.3. Stars with shallow convection zones: $r_0 = 0.775$

With differential rotation defined by $a = 0.1$, the onset of dynamo action is at $R_\alpha \approx -2.0$. For $R_\alpha \sim -7.5$ to $-10.0$ fragmentation occurs, and at $R_\alpha \sim -15$ the regime becomes irregular.

With $a = 0.01$, the marginal $R_\alpha \approx -16.0$, and the solutions start to become noisy, and eventually become fragmented, at $|R_\alpha| \gtrsim 20.0$.

As examples we show in Figs. 8 and 9 comparable results to those given previously, for $a = 0.1$ and $a = 0.01$ respectively.

### 3.4. Mean rotation profiles

An important question is the way and the extent to which the magnetic field modifies the rotation profile. In order to quantify the changes in the rotation profile produced by this interaction, we studied the mean changes in $\Omega(r, \theta)$ (normalised by $\Omega_0$) as a function of $r$ and $\theta$ over a cycle. In particular we considered the quantity

$$\langle \Delta\Omega(\Omega_0, r, \theta, t)\rangle = \underbrace{\text{Average}}_{\text{time}} \left[ \frac{|\Omega(r, \theta, t) - \Omega_0(r, \theta)|}{\Omega_0(r, \theta)} \right] \quad (5)$$

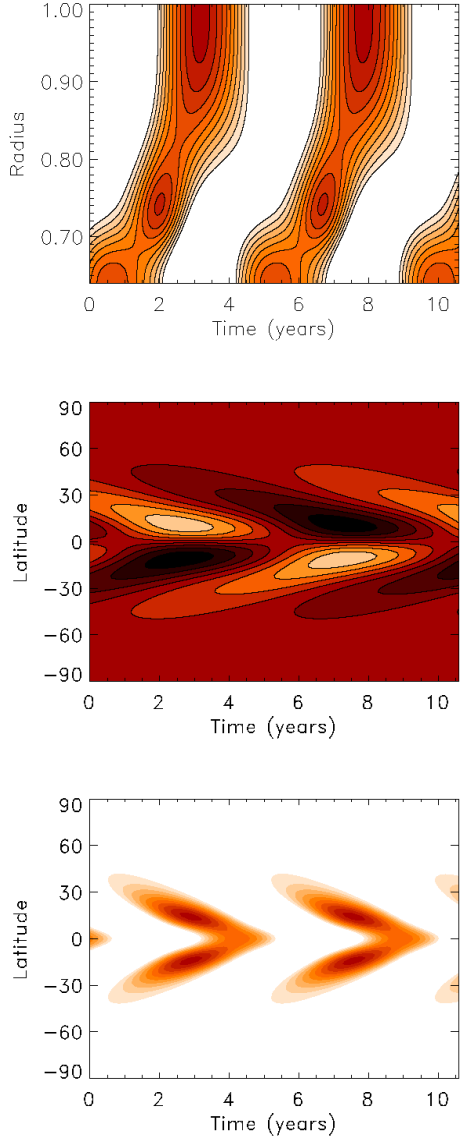

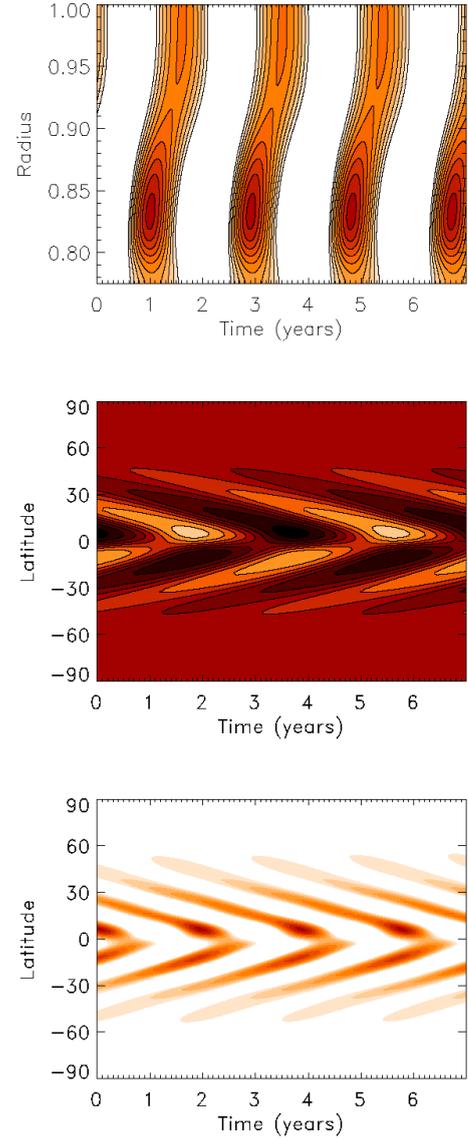

**Fig. 7.** As in Fig. 2 for a dynamo region with $r_0 = 0.64$, $a = 0.01$ and $R_\alpha = -12$. Note the penetration of the torsional oscillations all the way down to the bottom of the dynamo region, and also the localisation of the torsional oscillations and the magnetic field near the equator.

**Fig. 8.** As in Fig. 2 for a dynamo region with $r_0 = 0.775$, $a = 0.1$ and $R_\alpha = -5$. Note the penetration of the torsional oscillations all the way down to the bottom of the dynamo region, and also the localisation of the torsional oscillations and the magnetic field near the equator.

We note that this is different from the torsional oscillation amplitude, as is commonly defined, given by

$$\underbrace{\text{Average}}_{\text{time}} |\Omega(r, \theta, t) - \langle\Omega(r, \theta, t)\rangle| \quad (6)$$

An example of the contours of $\langle\Delta\Omega\rangle$ normalised by $\Omega_0$, defined by Eq. (6) is given in Fig. 12 for a solar-type model with $r_0 = 0.64$, $R_\alpha = -2.0$ and $a = 0.1$.

Briefly these results show that in all cases, the effects are mainly concentrated near the equator and the bottom of the dynamo region. The strongest surface effects (i.e. largest $|\langle\Delta\Omega\rangle|$) are seen for the stars with the deepest dynamo region ($r_0 = 0.2$). The general shape of the contours is similar in all three cases, with a band of strong contours starting near $r = r_0$ at the equator and coming to the surface in mid latitudes.

To give quantitative estimates of $\langle\Delta\Omega\rangle$, we show in Fig. 10 its behaviour as a function of $R_\alpha$ for models with different depths of the dynamo region. The maximum values of $\langle\Delta\Omega\rangle$ for the cases with $a = 0.1$ are 2.3%, 2.7% and 2.1% for $r_0 = 0.2$,

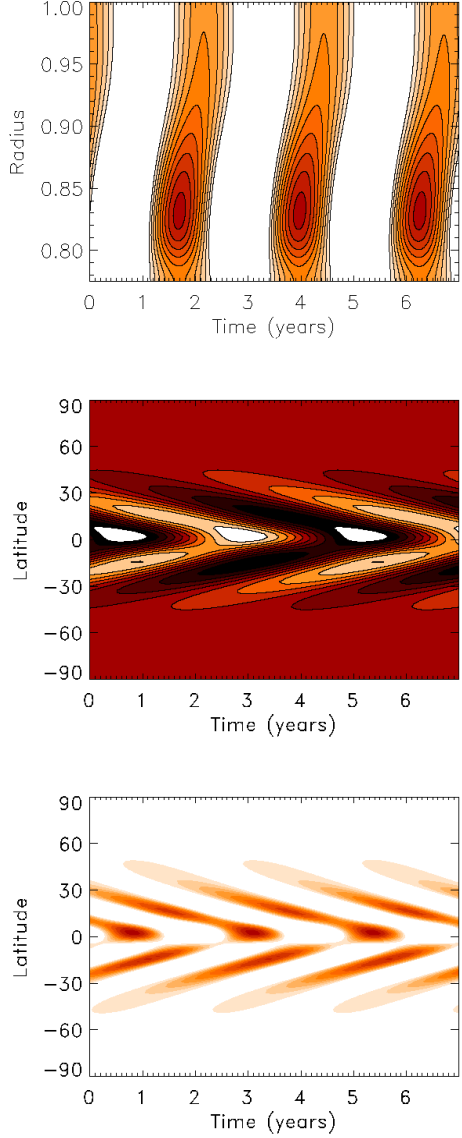

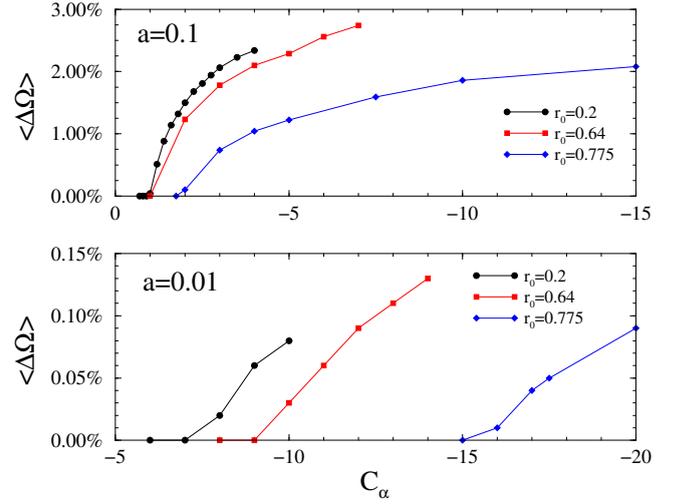

**Fig. 10.** The behaviour of $\langle \Delta \Omega \rangle$ as a function of $R_\alpha$ for models with different depths of the dynamo region.

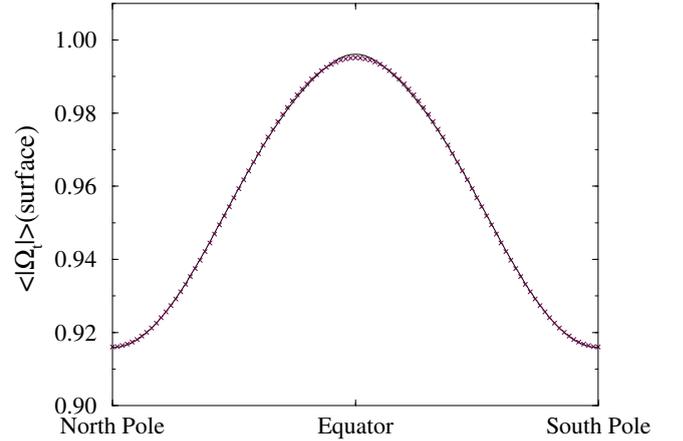

**Fig. 11.** The fitting of our rotation profile by the parameterization (7). The model parameters used are $r_0 = 0.2, R_\alpha = -1.40$ and $a = 0.1$. The continuous curve gives the values of $<|\Omega_t|>$ at the surface, whereas the crosses correspond to the interpolation $y = a_0 + a_1 \sin^2 \theta + a_2 \sin^4 \theta$ – see text.

**Fig. 9.** As in Fig.2 for a dynamo region with $r_0 = 0.775, a = 0.01$ and $R_\alpha = -17$. Note the penetration of the torsional oscillations all the way down to the bottom of the dynamo region, and also the localisation of the torsional oscillations and the magnetic field near the equator.

$r_0 = 0.64$ and $r_0 = 0.775$ respectively, and when $a = 0.01$ they are 0.08%, 0.13% and 0.09% for $r_0 = 0.2$, $r_0 = 0.64$ and $r_0 = 0.775$ respectively.

As can be seen, the largest amplitudes of the deviations are of the order of few percent. The observation that the values of $\Delta\Omega$ are small is interesting as it implies that the underlying imposed angular velocity $\Omega_0$ is not very strongly distorted by its interaction with the magnetic field.

To estimate the effects of the magnetic field on the observable surface rotation profile, we also studied how well the mean surface rotation law can be fitted by the rotation law usually employed by observers in fitting their data

$$\Omega = \Omega_*(a_0 + a_1 \sin^2 \theta + a_2 \sin^4 \theta) \qquad (7)$$

where $a_0(\sim 1), a_1$ and $a_2$ are convenient parameterisations of the data and in our case the rotation profile. The results are presented in Fig. 11, for model parameters $r_0 = 0.2, R_\alpha = -1.40$ and $a = 0.1$. The values of the fitted parameters were found to be $a_0 = 0.916069, a_1 = 0.0840602$ and $a_2 = -0.00500258$. This figure shows that the changes to the underlying differential rotation $\Omega_0$ due to the torsional oscillations are small.

Of course, in general both model and observed rotation laws vary with time, but such variations are only now beginning to be determined observationally. This aspect is addressed in Sect. 4.

### 3.5. Anti–solar surface rotation

We also investigated the effects of changing the sign of $a$ in the definition of the underlying rotation profile, whilst keeping $R_\alpha < 0$. This is of potential relevance since there are hints that such behaviour may occur in a minority of stars (e.g. Messina & Guinan 2003).

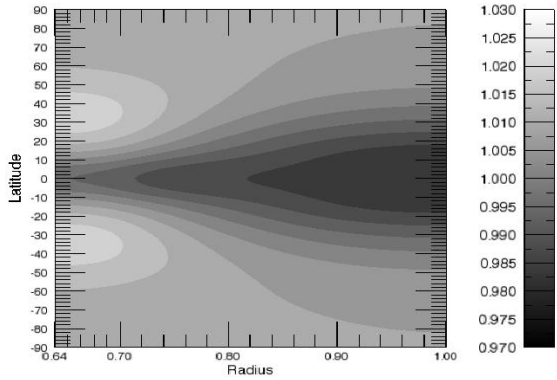

**Fig. 12.** The rectangular contours of the average $\Omega(r,\theta)$ normalized by $\Omega_0$, defined by (6) for a solar-like model with $r_0 = 0.64$, $R_\alpha = -2$ and $a = 0.1$.

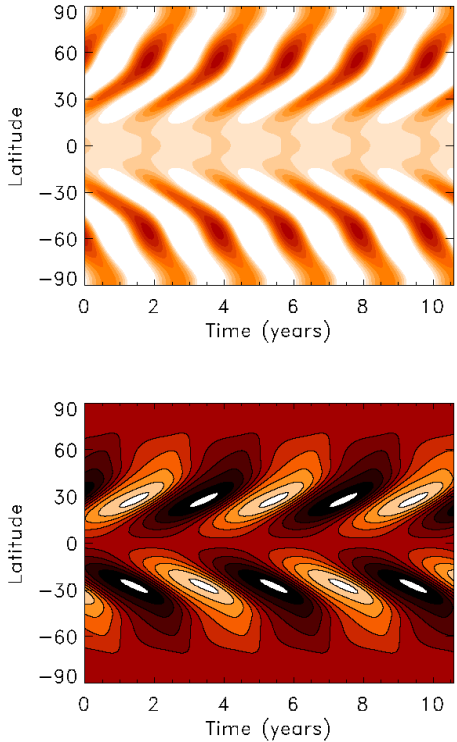

**Fig. 13.** The butterfly diagrams of the near-surface magnetic field and torsional oscillations for a dynamo region with $r_0 = 0.64$, $a = -0.1$ and $R_\alpha = -3$.

We show a typical case in Fig. 13. Unsurprisingly perhaps, the major effect is that now the sense of migration of the torsional oscillations (and the magnetic field) is reversed.

## 4. Comparison with observations

For a small number of objects, notably AB Dor and LQ Hyd, observations are of sufficiently high quality to allow determination of not only the mean surface rotation law $\Omega_s$, but also its variation with time. That is

$$\Omega_s(\theta, t) = \Omega_{\text{pol}}( \quad 1 + a_{*01} \cos(\omega t + \delta) + \quad (8)$$
$$(a_{*10} + a_{*11} \cos(\omega t)) \sin^2 \theta),$$

where $\Omega_{\text{pol}}$ is the mean polar angular velocity and the quantities $a_{*01}, a_{*10}$ and $a_{*11}$ are known. (Note that here, as in all of this paper, $\theta$ is polar angle, not latitude, and so this law takes a slightly different form from that often presented by observers.) Empirically we find that in our solutions, the phase-shift $\delta \approx \pi/2$, and that $|a_{*01}| \ll |a_{*11}|$, so in this first, order of magnitude analysis, we ignore $a_{*01}$. An estimate of the appropriate value of our parameter $a$ (Eq. 3) to represent such a rotation is given by $a_{*10}$: however this can be regarded only as a first approximation, as it is clear from our analysis in Sect. 3.4 that in the finite amplitude state, the mean of the coefficient of $\sin^2 \theta$ differs slightly from the prescribed $a$. In our models dynamo action is determined by the product $R_\omega a/(1 + a) = R_\omega^*$, and for given $R$ and $\eta_0$, it is determined by $\Omega_0 a/(1 + a)$. Thus we can use our computations for different parameters $a_*$, $\Omega_{0*}$ provided that

$$\frac{a}{(1+a)} = \frac{\Omega_{0*}}{\Omega_0} \frac{a_*}{(1+a^*)}. \quad (9)$$

We are then in a position to make a first, order of magnitude, comparison with the observed behaviour of AB Dor and LQ Hyd. Of course, the value of $\eta_0$ is really quite uncertain, and at this stage we can do no more than see if the values are at all plausible.

### 4.1. AB Dor

Observations (e.g. Collier Cameron & Donati 2002, Donati *et al.* 2003b) suggest that in Eq. (9), $a_{*10} \sim 0.007$, $a_{*11} \sim 0.0015$, and $\Omega_{\text{pol}} \sim 55\Omega_\odot$. Thus if we estimate the appropriate zero order rotation law very approximately to correspond to $a \sim a_1 \sim 0.007$, then with our canonical values for $\eta_0$ and $\Omega_{\text{pol}} = 10\Omega_\odot$, and using relation (9), the corresponding dynamo calculation has $a \approx 0.04$. Our closest value is $a = 0.05$. However such an estimate for $a$ should not be taken too literally, given the arbitrariness in choice of $\eta_0$. The most relevant parameter is probably $a_{*11}/a_{*10} = \xi$ say. Thus for AB Dor $\xi$ lies in the range $0.1 - 0.2$.

For a dynamo region with $r_0 = 0.2$ and $R_\alpha = -2.0$, our solution with $a = 0.05$ gave $a_{*10} \approx \langle a_1 \rangle = 0.048$, $a_{*11} \approx \frac{1}{2}[\max(a_1) - \min(a_1)] = 7.05 \times 10^{-5}$, i.e. $\xi \approx 0.0015$. When $a = 0.1$, $a_{*10} = 0.078, a_{*11} = 0.0025, \xi = 0.03$, and when $a = 0.2$, $a_{*10} = 0.17$, $a_{*11} = 0.004$, so that $\xi = 0.025$. For the case $r_0 = 0.64$ and $R_\alpha = -3.0$, we found $a_{*10} \approx \langle a_1 \rangle = 0.05$ and $a_{*11} \approx 0.0015$, i.e. $\xi \approx 0.03$. For $a = 0.1$, we find $a_{*10} = 0.092, a_{*11} = 0.011, \xi = 0.12$, and when $a = 0.2$ we obtained $a_{*10} = 0.18$, $a_{*11} = 0.026$, $\xi = 0.14$. When $|R_\alpha|$ is increased above these values, solutions lose their regular, coherent behaviour, although the departure from the zero-order angular velocity distribution can become larger.

*4.2. LQ Hyd*

For this star, the surface angular velocity appears to vary much more dramatically, with $a_{*10} \sim 0.25 \sim a_{*11}, \xi \approx 1$ (i.e. at times the surface rotation is approximately solid body), and now $\Omega_{pol} \sim 19\Omega_\odot$ (Donati *et al.* 2003b). By the arguments given in Sect. 4.1, we again estimate a nominal value $a = 0.05$. Thus the dimensionless results are as given above for AB Dor, i.e. when $r_0 = 0.2$, $\xi \approx 0.0015, 0.03$, and when $r_0 = 0.64$ then $\xi \approx 0.03, 0.12$, respectively.

## 5. Summary

We have made a detailed study of the variations in magnetic field and angular velocity that might be expected to occur in the convection zones of rapidly rotating cool stars, using a mean field dynamo model in which the only nonlinearity is the action of the azimuthal component of the Lorentz force of the dynamo generated magnetic field on the stellar angular velocity. This dynamo model, when applied to the Sun, yields results that have an encouraging agreement with the results from the helioseismic observations. Given the current difficulties in resolving the precise nature of such variations a theoretical understanding of the range of possible modes of behaviour is crucial in interpretation the observations.

We considered three families of models with different depths of dynamo-active regions, with bases at fractional radii $r_0 = 0.2, 0.64$ and $0.775$. For moderately critical regimes we find torsional oscillations to be present, qualitatively similar to those found in the Sun, that in all cases penetrate all the way down to the bottom of the convection zone/dynamo-active region. [1] In some cases, the surface amplitudes are small, and the oscillations are concentrated deep down. For larger values of $|R_\alpha|$ we found fragmentation in some cases as well as other dynamical modes of behaviour, including quasi-periodicity and irregularity.

We have found that the values of $\langle\Delta\Omega\rangle$, the averaged relative deviations in $\Omega_0$, increase with $|R_\alpha|$ as well as with increasing values of the strength of the differential rotation parameter $a$ or of $R_\omega^*$. The largest values of $\langle\Delta\Omega\rangle$ (obtained for $a = 0.1$) are of the order of a few percent for coherent non-chaotic regimes. The fact that these are relatively small is interesting as it implies that the original angular velocity law is not very distorted – quite modest changes from $\Omega_0$ are enough to self-limit dynamo action (cf. also Moss & Brooke 2000). In this connection, we have also shown that the rotation profile can be well represented by a rotation law of the form (7).

When we make a preliminary comparison with the observed behaviour of the late-type dwarfs AB Dor and LQ Hyd (Sect. 4), we can find models with variations in surface differential rotation that are not inconsistent with those observed for these stars, although we cannot produce anything as extreme as the episodes of almost uniform surface rotation as reported for LQ Hyd. In fact, our model cannot distinguish between the representations of these stars – this points to omissions in our modelling. For example, we do not directly represent absolute differences in angular velocity between different models, as these models depend only on differential rotation. *Inter alia*, the absolute value of the angular velocity might be expected to affect the structure of the alpha-tensor (here reduced to a scalar). However, we are encouraged; our very simple model is untuned, and there are a number of plausible modifications that could be made to it.

Our results all show butterfly diagrams which are mainly concentrated around the equator (with dispersion depending upon the parameters of the model. This differs from the results recently obtained by Bushby (2003) who, with a different $\alpha$-quenched dynamo model, used an $\alpha$-profile concentrated near the tachocline together with a quasi-cylindrical rotation profile with $\Omega \propto r\sin\theta$ (rather than $\propto r^2\sin^2\theta$, as suggested by observations and used in our model). We made a comparison, using our code with $\alpha$-quenching implemented together with his $\alpha$ and $\Omega$ profiles, and found agreement with his results. We, however, did find that Bushby's results are somewhat fragile, in the sense that there seem to be two distinct regimes distinguished by the strength of the differential rotation parameter. In particular we found that increasing this parameter from 0.0046 (as used by Bushby) to 0.01 changes the behaviour of the butterfly diagrams from a pole to an equator dominated regime. That choice (0.0046) was motivated by a desire in an $\alpha$-quenched model with unchanging angular velocity, to have a rotation law that exerted zero total rotational stress on the interior region. We also find that these results are fragile with respect to the choice of the $\Omega$ profile, specifically by changing $\Omega \propto r\sin\theta$ to $r^2\sin^2\theta$, as well as to the radial distribution of $\alpha$ noted by Bushby.

---

[1] The most recent inversions also show evidence for the presence of oscillations down to the deeper layers of the convection zone (Vorontsov *et al.* 2002, Antia & Basu 2004), but the precise dynamical nature of these regimes cannot be determined at present, given the current length of the helioseismological data set and the presence of noise. We have however shown, using a detailed comparison between model predictions and helioseismological inversions, that given the expected noise levels in the helioseismological data, our theoretical predictions are compatible with observations (Vorontsov *et al.* 2003). Our models also predict a polar branch of the torsional oscillations, and this feature has also been confirmed.